\documentclass[%
 aip,
 jmp,%
 amsmath,amssymb,
 reprint,%
]{revtex4-1}

\usepackage{graphicx}
\usepackage{dcolumn}
\usepackage{bm}
\usepackage{graphicx}
\usepackage{dcolumn}
\usepackage{bm}
\usepackage{units}
\usepackage{xspace}
\usepackage{textcomp} 
\usepackage{amsmath,amsfonts,amssymb}
\usepackage{setspace}
\usepackage{tocloft}
\usepackage{placeins}
\usepackage{txfonts}
\usepackage{float}
\usepackage{natbib}

\newcommand{\dose}[1]{\unit[#1]{$\text{H}^{+}\text{cm}^{-2}$}\xspace}
\newcommand{\diffusivity}[1]{\unit[#1]{$\text{cm}^{2}\text{s}^{-1}$}\xspace}
\newcommand{\Celsius}[1]{\unit[#1]{\ensuremath{^{\circ}}C}\xspace}
\newcommand{\MeV}[1]{\unit[#1]{MeV}\xspace}

\newcommand{\eV}[1]{\unit[#1]{eV}\xspace}
\newcommand{\mum}[1]{\unit[#1]{\textmu m}\xspace}
\newcommand{\nm}[1]{\unit[#1]{nm}\xspace}
\newcommand{\sqcm}[1]{\unit[#1]\ensuremath{{\text{cm}^{2}}}\xspace}
\newcommand{\minute}[1]{\unit[#1]{min}\xspace}
\newcommand{\percent}[1]{\unit[#1]{\%}\xspace}
\newcommand{\hour}[1]{\unit[#1]{h}\xspace}

\newcommand{\Kelvin}[1]{\unit[#1]{K}\xspace}
\newcommand{\heatingrate}[1]{\unit[#1]{K/s}\xspace}
\newcommand{\concentration}[1]{\unit[#1]{$\text{cm}^{-3}$}\xspace}

\begin{document}

\preprint{AIP/123-QED}

\title{Diffusion of Hydrogen in Proton Implanted Silicon: Dependence on the Hydrogen Concentration}  

\author{Martin Faccinelli}
	\affiliation{Institute of Solid State Physics, Graz University of Technology}%
\author{Stefan Kirnstoetter}%
	\affiliation{Institute of Solid State Physics, Graz University of Technology}%
\author{Moriz Jelinek}
	\affiliation{Infineon Technologies Austria AG}
\author{Thomas Wuebben}
	\affiliation{Infineon Technologies Austria AG}
\author{Johannes G. Laven}
	\affiliation{Infineon Technologies AG}
\author{Hans-Joachim Schulze}
	\affiliation{Infineon Technologies AG}
\author{Peter Hadley}
	\affiliation{Institute of Solid State Physics, Graz University of Technology}
\date{\today}
\begin{abstract}
The reported diffusion constants for hydrogen in silicon vary over six orders of magnitude. This spread in measured values is caused by the different concentrations of defects in the silicon that has been studied. Hydrogen diffusion is slowed down as it interacts with impurities. By changing the material properties such as the crystallinity, doping type and impurity concentrations, the diffusivity of hydrogen can be changed by several orders of magnitude. In this study the influence of the hydrogen concentration on the temperature dependence of the diffusion in high energy proton implanted silicon is investigated. We show that the Arrhenius parameters, which describe this temperature dependence decrease with increasing hydrogen concentration. We propose a model where the relevant defects that mediate hydrogen diffusion become saturated with hydrogen at high concentrations. When the defects that provide hydrogen with the lowest energy positions in the lattice are saturated, hydrogen resides at energetically less favorable positions and this increases the diffusion of hydrogen through the crystal. Furthermore, we present a survey of different studies on the diffusion of hydrogen. We observed a correlation of the Arrhenius parameters calculated in those studies, leading to a modification of the Arrhenius equation for the diffusion of hydrogen in silicon.  

\end{abstract}
\pacs{Valid PACS appear here}
\maketitle

\section{\label{sec:Introduction}Introduction}
For an accurate description of the properties of silicon, knowledge of the diffusion behaviour of impurities is crucial\cite{Pichler2004,Pearton2013}. The diffusion of hydrogen in silicon has been studied extensively.\cite{vanWieringen1956,Ichimiya1968,Mogro-Campero1985,Pearton1985,Kazmerski1985,Seager1988,Herrero1990,Rizk1991,Newman1991,Johnson1991,Sopori1992,Panzarini1994,Huang2004,Laven2013} The experimental data obtained in the different studies is difficult to interpret because the diffusion constants measured in these studies vary up to more than six orders of magnitude at a given temperature. 

Phenomenologically, the propagation of hydrogen can be described by Fick's second law of diffusion
\begin{equation}
\label{equ:Ficks2}
\dfrac{\delta c}{\delta t}=D\dfrac{\delta^2c}{\delta x^2}\textrm{,}
\end{equation}
where, $D$ is the diffusivity and $c$ is the concentration of the diffusing hydrogen. The temperature dependence of the diffusion is typically observed to follow an Arrhenius law
\begin{equation}
\label{equ:Arrhenius}
D=D_0 e^{-\frac{E_A}{k_BT}},
\end{equation}
where $D_0$ is a pre-factor, $E_A$ is the activation energy and $k_B$ is Boltzmann's constant.

In the presence of defects, the migration of hydrogen is slowed down as the hydrogen reacts with defects such as dopants, vacancies, and interstitials.\cite{Herrero1990} These hydrogen-defect complexes that form have a dissociation rate so the hydrogen is later released whereupon it diffuses and reacts again. The slowed-down diffusion of hydrogen is still described by equation~\ref{equ:Ficks2} but is characterized by a defect mediated, effective diffusivity. The type of defects present and their concentrations have a strong influence on the diffusion of hydrogen.

In this study, the influence of the implantation dose on the effective diffusivity in high energy proton implanted silicon is investigated. The implantation of protons into silicon followed by a subsequent annealing step can generate a variety of different defect complexes. Some of these complexes act as recombination centers and are used to fine-tune the lifetime of minority charge carriers~\cite{Hazdra2001}. Other defect complexes can grow to extended defects and are used for cleaving thin layers off a silicon wafer in the so-called "Smart-Cut" process~\cite{Romani1990}. There are even other defect complexes which have energy states close to the conduction band, and hence, act as donors~\cite{Zohta1971,Wondrak1986}. The kind of hydrogen related defect complexes that are formed, depends on the implanted hydrogen dose, the annealing treatment, and the defects already present in the material. These other defects can be intrinsic defects such as vacancies and interstitials, which are formed during the proton implantation, or impurities like oxygen, carbon or dopants. 

In the proton implantation process, a high concentration of hydrogen is introduced into a narrow region around the implantation depth. During the anneal, the hydrogen diffuses away from this region towards the surfaces of the material. On its way through the proton irradiated region it forms hydrogen related defect complexes. The spatial distribution of electrically active complexes can be measured using techniques such as spreading resistance profiling (SRP)~\cite{Mazur1966} or Capacitance-Voltage measurements (CV)~\cite{Zohta1971}. 

\section{\label{sec:Experimental}Experimental Procedure}

\begin{figure}[tbp]%
 \includegraphics[width=\linewidth]{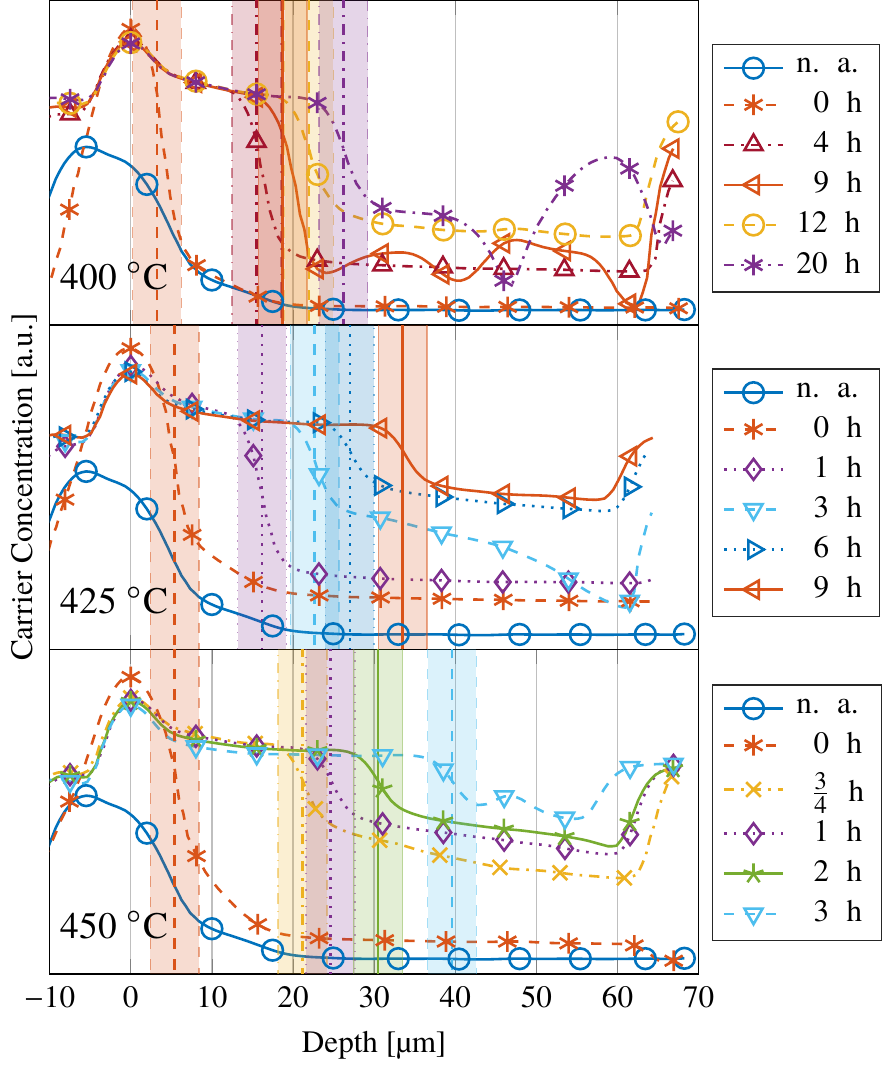}
\caption{Charge carrier concentration profiles of sample S1, after an annealing at temperatures from \Celsius{400} to \Celsius{450} for various annealing times. The extracted diffusion length $L_D$ is indicated as a vertical line. The shaded areas represent an error of \mum{$\pm3$}. The charge carrier concentration is plotted on a logarithmic scale.
 }
\label{fig:SRP}
\end{figure}

One float-zone (FZ) and two magnetic Czochralski (m:Cz) silicon wafers were implanted with \MeV{2.5} and \MeV{4} protons at doses ranging from \dose{$10^{14}$} to \dose{$10^{16}$}. The wafers were cut into pieces of \sqcm{$1\times1$} and annealed at temperatures from \Celsius{400} to \Celsius{500} in an Anton Paar DHS 1100 heating stage under N$_2$ atmosphere. The annealing time was varied from \minute{15} to \hour{20}. Table~\ref{tab:samples} lists the samples that were investigated. The distribution of electrically active defects generated during the anneal was determined by measuring the charge carrier concentration profiles with SRP measurements at room temperature. In this method, the sample is ground at a small angle and the resistance between two metal tips, which are brought into contact with the sample, is measured as a function of the position along the bevel. As described in reference~\citenum{Mazur1966}, the resulting resistance profile is then converted into a charge carrier concentration (CCC)-profile.
\begin{table}[tbp]
\begin{ruledtabular}
\begin{tabular}{c c c c c}
& Material & O [\concentration{}]   & Energy [MeV] &  Dose [\dose{}]  \\
  \colrule
S1 & \textit{p}-m:Cz& $< 4 \times 10^{17}$ & {2.5}	& $5\times10^{14}$ \\
S2 & \textit{p}-m:Cz& $< 4 \times 10^{17}$ & {2.5} 	& $1\times10^{15}$ \\
S3 & \textit{p}-m:Cz& $< 4 \times 10^{17}$ & {2.5} 	& $5\times10^{15}$ \\
S4 & \textit{p}-m:Cz& $< 4 \times 10^{17}$ & {2.5} 	& $1\times10^{16}$ \\
S5 & \textit{p}-m:Cz& $< 4 \times 10^{17}$ & {4.0} 	& $1\times10^{14}$ \\
S6 & \textit{p}-m:Cz& $< 4 \times 10^{17}$ & {4.0} 	& $1\times10^{15}$ \\
S7 & \textit{p}-m:Cz& $< 2 \times 10^{17}$ & {2.5} 	& $1\times10^{14}$ \\
S8 & \textit{n}-Fz	& $< 1 \times 10^{17}$ & {4.0} 	& $1\times10^{14}$ \\
\end{tabular}
\end{ruledtabular}
\caption
{\label{tab:samples}
List of investigated samples including material and doping type, oxygen concentration and implantation parameters.}
\end{table}
Figure~\ref{fig:SRP} shows a set of CCC-profiles of pieces of sample S1, annealed at different temperatures for different annealing times. Here, the annealing time was the time between reaching the plateau temperature and the beginning of the cooling. The heating and cooling rates were roughly \heatingrate{1}. The data also includes CCC-profiles before the annealing (n. a.) and CCC-profiles where the sample was heated to the plateau temperature and immediately cooled down to room temperature (\hour{0}). It can be observed, that a region with an increased concentration of charge carriers is formed, which expands, as the annealing time is increased. This expansion can be described by the expression  
\begin{equation}
\label{equ:diffusion_length}
L_D=L_0+\sqrt[]{4Dt}\textrm{,}
\end{equation}
where $L_D$ is the diffusion length and $L_0$ is the depth from which the diffusion begins.

\begin{figure}[tbp]%
 \includegraphics[width=\linewidth]{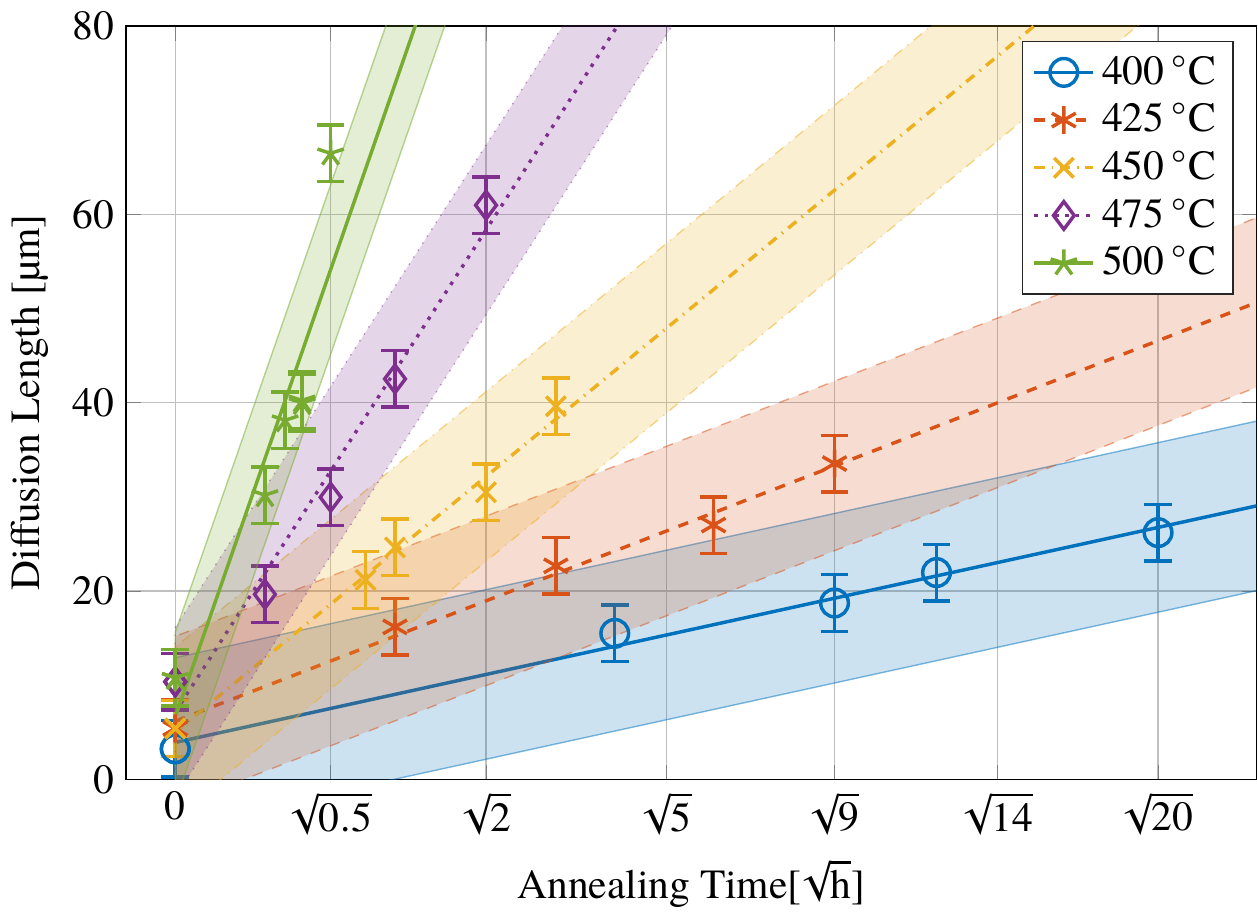}
\caption{Diffusion length $L_D$ of hydrogen in S1 plotted as a function of the square root of the annealing time for annealing temperatures from \Celsius{400} to \Celsius{500}. The straight lines represent the linear fit through the measurement points, including the $3\sigma$-confidence interval, indicated by the shaded areas.
 }
\label{fig:diffusion_length}
\end{figure}

If SRP is used for measuring the effective diffusivity of hydrogen, the diffusion length is usually extracted from the position of a $pn$-junction formed in the material~\cite{Huang2004,Laven2013}. Due to the low doping concentrations in the investigated samples (below \concentration{$5\times10^{13}$}), the formation of this junction was not always observed. Hence, the diffusion length $L_D$ of the effective hydrogen diffusion was extracted as the position of the steepest decay of the CCC at the edge of the region with increased charge carrier concentration. These positions are indicated by vertical lines in figure~\ref{fig:SRP}. The diffusion length extracted after heating the sample and cooling it down again~(\hour{0}) was used as $L_0$. The shaded areas in the figure represent the range of the error of $L_D$. The error was estimated in such a way that the measured values of $L_D$ would lie inside the $3\sigma$-confidence interval, represented by the shaded area in figure~\ref{fig:diffusion_length}. This approximated error corresponds to a measurement error of \mum{$\pm3$}. Figure~\ref{fig:diffusion_length} shows the diffusion length of hydrogen in sample S1 as a function of the square root of the annealing time for different annealing temperatures. The effective diffusivity $D$ is extracted from the slope according to equation \ref{equ:diffusion_length}. The measurement error of the diffusion length is indicated by error bars in figure~\ref{fig:diffusion_length}. The straight lines represent the linear fit through the measurement points. The shaded areas indicate the $3\sigma$-confidence interval. In table~\ref{tab:diffusivities} the effective hydrogen diffusivities in all samples and for all anneals are listed.
\begin{table*}[tp]
\begin{ruledtabular}
\begin{tabular}{c  c c c c c c}
  &\Celsius{400} &\Celsius{425} &\Celsius{450} &\Celsius{475} &\Celsius{500} \\
\hline
S1	& $1.8 \times10^{-11}\pm\percent{41}$ & $5.8 \times10^{-11}\pm\percent{40}$ &	$2.5 \times10^{-10}\pm\percent{27}$ &	$8.9 \times10^{-10}\pm\percent{21}$ & $3.2 \times10^{-09}\pm\percent{29}$ \\
S2	& $2.1 \times10^{-11}\pm\percent{37}$ & $5.1 \times10^{-11}\pm\percent{42}$ &	$1.3 \times10^{-10}\pm\percent{36}$ &	$8.0 \times10^{-10}\pm\percent{23}$ & $1.9 \times10^{-09}\pm\percent{25}$ \\
S3	& $1.1 \times10^{-10}\pm\percent{20}$ & $2.7 \times10^{-10}\pm\percent{25}$ &	$9.1 \times10^{-10}\pm\percent{31}$ &	$2.4 \times10^{-09}\pm\percent{19}$ & $3.9 \times10^{-09}\pm\percent{22}$ \\
S4	& $5.4 \times10^{-11}\pm\percent{28}$ & $1.4 \times10^{-10}\pm\percent{30}$ &	$1.6 \times10^{-10}\pm\percent{21}$ &	$4.3 \times10^{-10}\pm\percent{39}$ & $7.7 \times10^{-10}\pm\percent{26}$ \\
S5	& $2.5 \times10^{-11}\pm\percent{55}$ &   					  & $4.8 \times10^{-10}\pm\percent{27}$ & 					     & 						\\
S6	& $1.7 \times10^{-11}\pm\percent{56}$ & 					  & $2.4 \times10^{-10}\pm\percent{20}$ & 						 & $3.3 \times10^{-09}\pm\percent{18}$ \\
S7	& $2.4 \times10^{-11}\pm\percent{59}$ & $9.0 \times10^{-11}\pm\percent{51}$ &	$2.7 \times10^{-10}\pm\percent{45}$ &	$1.1 \times10^{-09}\pm\percent{33}$ & $4.3 \times10^{-09}\pm\percent{27}$ \\
S8	& $8.4 \times10^{-11}\pm\percent{66}$ & 					  &	$8.2 \times10^{-10}\pm\percent{24}$ & 						 & 						\\
\end{tabular}
\end{ruledtabular}
\caption
{\label{tab:diffusivities}
Effective diffusivities of hydrogen in different materials at different temperatures in [$\text{cm}^{2}\text{s}^{-1}$].}
\end{table*}

\begin{figure}[tbp]%
 \includegraphics[width=\linewidth]{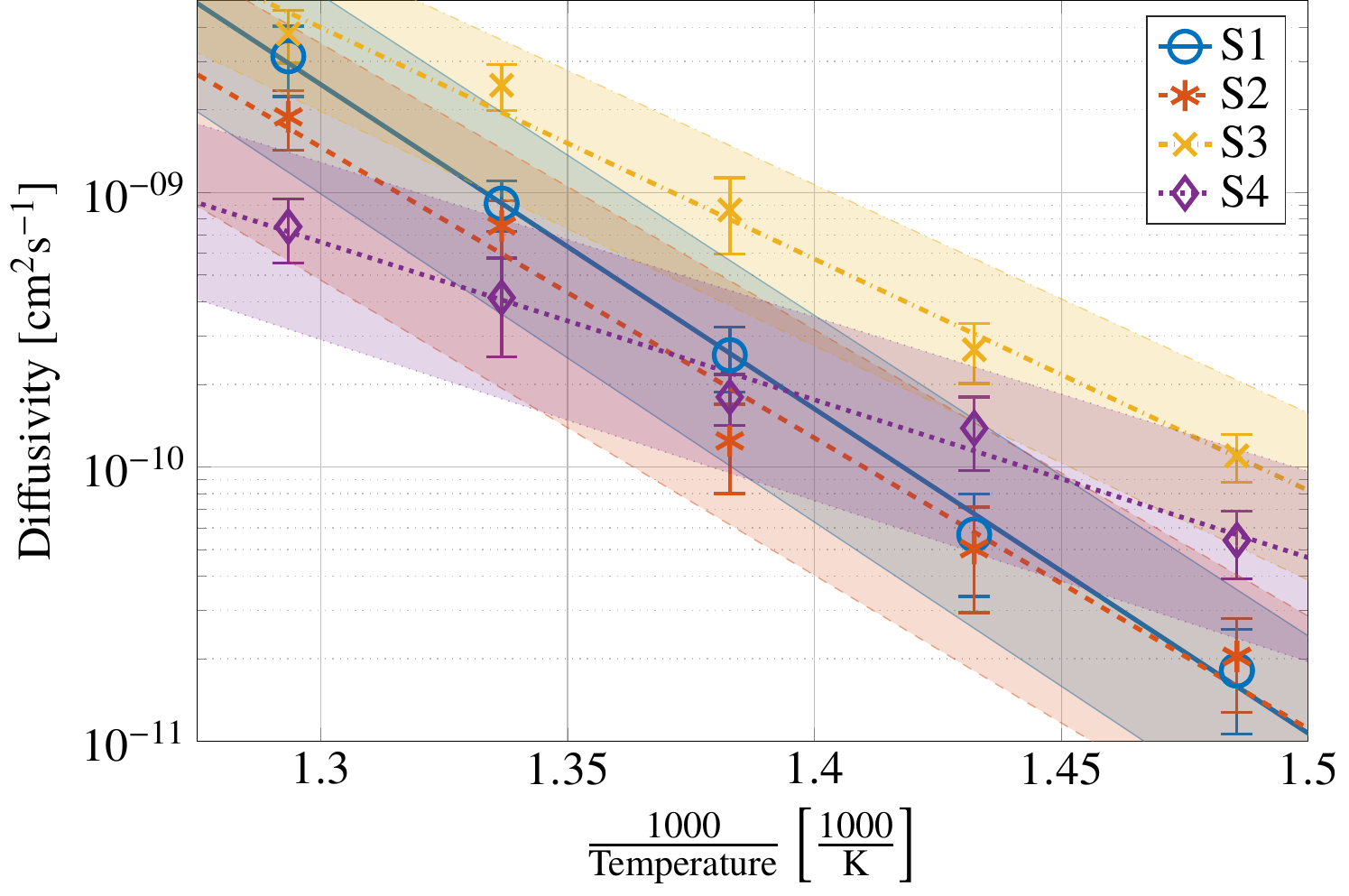}
\caption{Effective diffusivity of hydrogen in S1-S4 as a function of the inverse annealing temperature. The straight lines represent a fit through the measurement points, the shaded areas indicate the confidence interval.}
\label{fig:Arrhenius}
\end{figure}

The temperature dependence of the diffusivity is described by equation~\ref{equ:Arrhenius}. Diffusivities have been calculated for different temperatures from \Celsius{400} to \Celsius{500}. When the hydrogen diffusivity is plotted on a logarithmic scale as a function of the inverse temperature (see figure~\ref{fig:Arrhenius}), the Arrhenius parameters can be extracted. The intersection with the $y$-axis yields the pre-factor $D_0$. From the slope the activation energy $E_A$ can be extracted.
Figure~\ref{fig:Arrhenius} also includes the error of the effective diffusivity indicated as error bars. Linear fits through the data are shown as straight lines. The shaded areas represent the confidence interval of the diffusivity calculated from the errors of the extracted Arrhenius parameters. Here the temperature error of the annealing stage of \Kelvin{$\pm2$} is included.

\section{\label{sec:Results}Results}
\subsection{\label{sec:Arrhenius}Arrhenius Parameters vs. Implanted Proton Dose}
\begin{figure}[tbp]%
 \includegraphics[width=\linewidth]{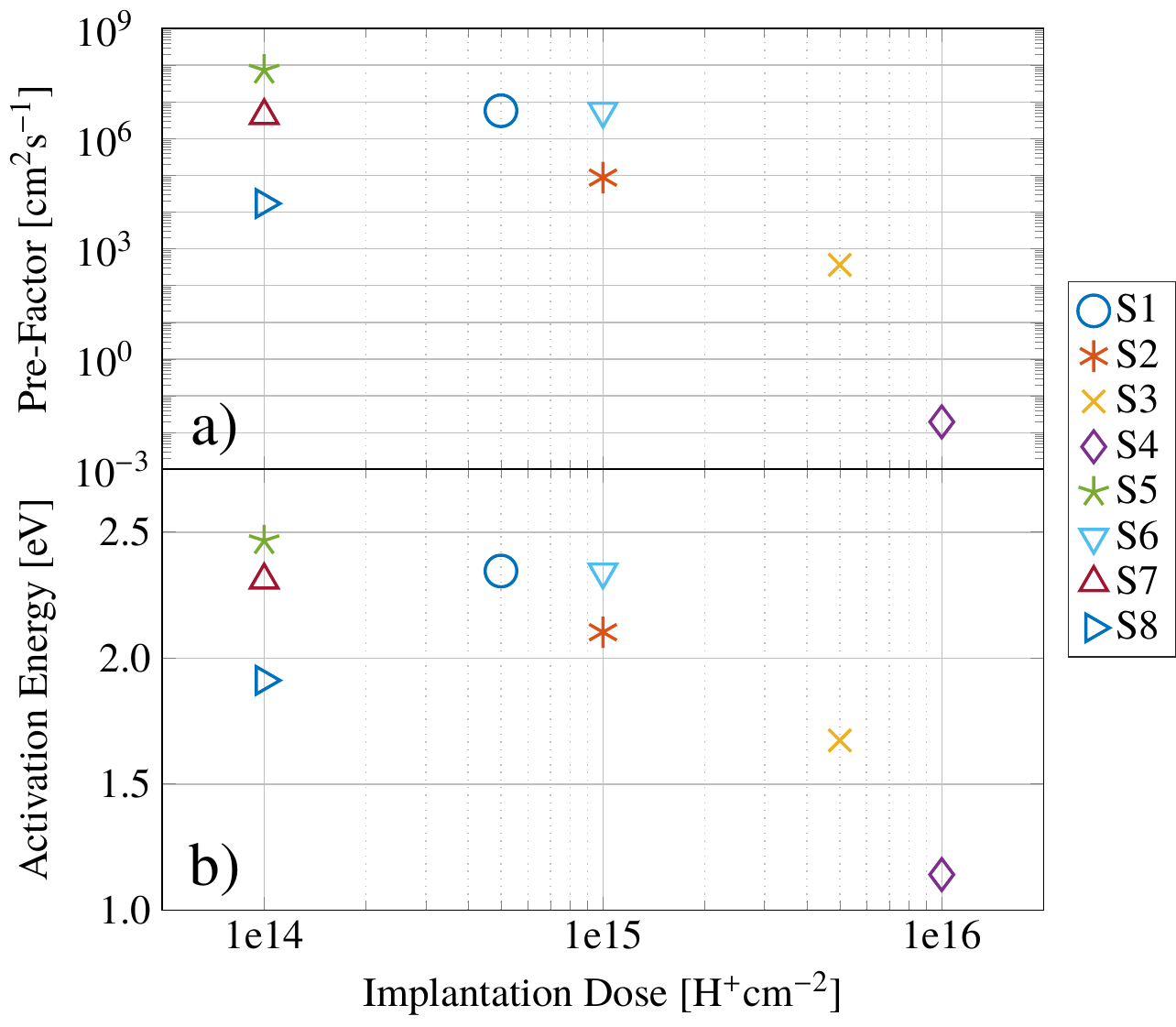}
\caption{Arrhenius parameters (a:$D_0$, b:$E_A$) of the samples S1-S8 as a function of the implanted proton dose.}
\label{fig:Arrhenius_parameters}
\end{figure}

In figure~\ref{fig:Arrhenius_parameters} the Arrhenius parameters of the hydrogen diffusivity of the investigated samples are plotted as a function of the implantation dose. In table~\ref{tab:Arrhenius_parameters} the Arrhenius parameters are listed. 

\begin{table}[tbp]
\begin{ruledtabular}
\begin{tabular}{c c c }
Sample&  $E_A$ [eV]&  $D_0$ [$\text{cm}^{2}\text{s}^{-1}$]\\
  \colrule
S1 & $2.34 \pm \percent{1}$ & $5.8 \times 10^{+6} \pm \percent{39}$  \\
S2 & $2.10 \pm \percent{2}$ & $8.7 \times 10^{+4} \pm \percent{47}$  \\
S3 & $1.67 \pm \percent{1}$ & $3.7 \times 10^{+2} \pm \percent{31}$  \\
S4 & $1.14 \pm \percent{2}$ & $2.0 \times 10^{-2} \pm \percent{37}$  \\
S5 & $2.47 \pm \percent{3}$ & $7.3 \times 10^{+7} \pm \percent{95}$  \\
S6 & $2.34 \pm \percent{2}$ & $5.6 \times 10^{+6} \pm \percent{60}$  \\
S7 & $2.31 \pm \percent{2}$ & $4.2 \times 10^{+6} \pm \percent{63}$  \\
S8 & $1.91 \pm \percent{3}$ & $1.7 \times 10^{+4} \pm \percent{86}$  \\
\end{tabular}
\end{ruledtabular}
\caption
{\label{tab:Arrhenius_parameters}
Activation energy $E_A$ and pre-factor $D_0$ of samples S1-S8 extracted from diffusivity measurements.}
\end{table}

We found (as shown in figure~\ref{fig:Arrhenius}) that at low temperatures the effective hydrogen diffusivity is higher in samples implanted with higher proton doses. On the other hand, the increase of the effective diffusivity with increasing temperature, described by the activation energy, decreases with increasing dose (see figure~\ref{fig:Arrhenius_parameters}). This leads to a higher effective hydrogen diffusivity in samples implanted with low doses at high temperatures. Furthermore, the results show a similar dependence of both the logarithm of the pre-factor and the activation energy on the implantation dose. For the samples S1-S4, where the same substrate material ($p$-m:Cz and [O]$<$\concentration{$4 \times 10^{17}$}) and implantation energy (\MeV{2.5}) has been used, the Arrhenius parameters both decrease with increasing implantation dose. If the implantation energy is varied (S2-\MeV{2.5} and S6-\MeV{4}), higher Arrhenius parameters were found if the sample was implanted at a higher energy. The results also show lower Arrhenius coefficients for $n$-FZ (S8) than for $p$-Cz (S5) material implanted under the same conditions (\MeV{4}, \dose{$10^{14}$}). 

\subsection{\label{sec:literature}Effective Hydrogen Diffusivity in Literature}

Most of the extracted Arrhenius parameters presented in section \ref{sec:Arrhenius} seem rather high when compared to literature, where reported activation energies for the effective diffusion of hydrogen lie between \eV{0.25} \cite{Kazmerski1985} and \eV{1.5}\cite{Huang2004}. We noticed that the activation energies and the Arrhenius pre-factors vary over a wider range than the diffusivity. This implies that the activation energy and the Arrhenius pre-factor are correlated with each other.
\begin{figure}[tbp]%
 \includegraphics[width=\linewidth]{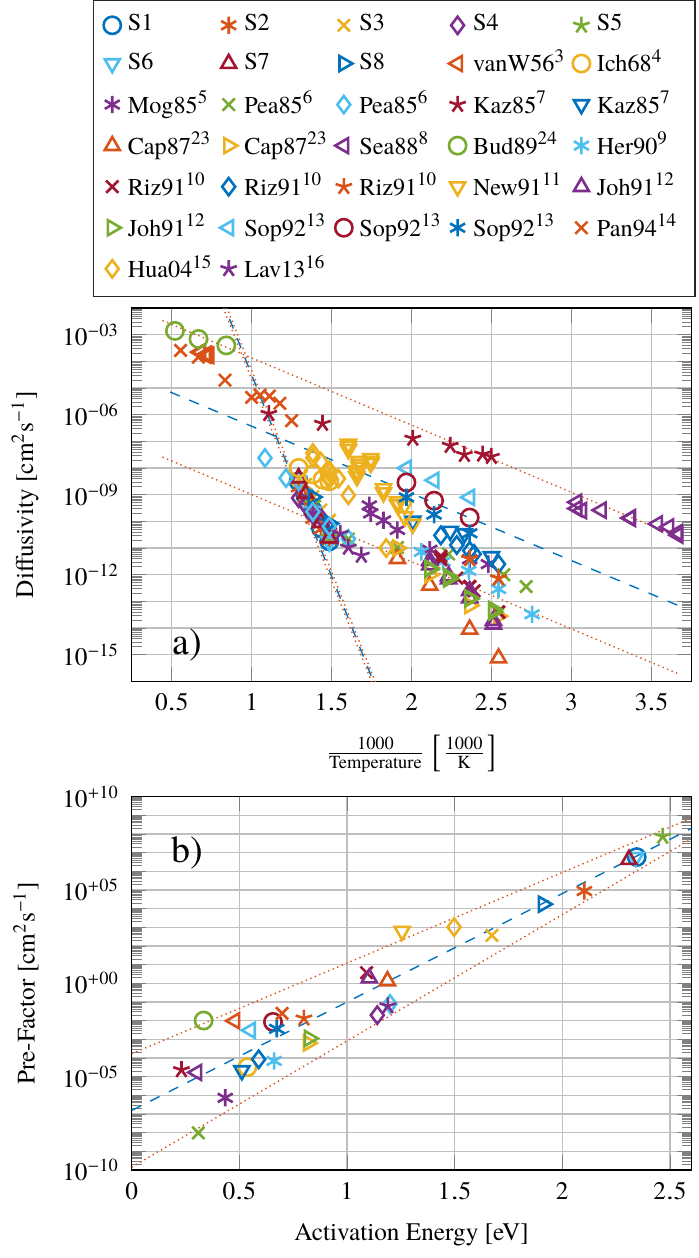}
\caption{a: Effective diffusivity of hydrogen as a function of the inverse temperature found in different studies. The dashed lines represent the characteristic of the effective hydrogen diffusivity, calculated for activation energies of \eV{0.5} and \eV{3}, using the best fits of $D_1$ and $E_\text{c}$. The dotted lines represent the characteristic of the effective hydrogen diffusivity calculated for activation energies of \eV{0.5} and \eV{3}, using the upper and lower limits of the parameters. b: Pre-Factor $D_0$ as a function of the activation energy $E_A$ for Arrhenius parameters calculated from the effective diffusivities plotted in figure~\ref{fig:Subplots}a. The dashed line is a linear fit through the data points with the slope $\frac{1}{E_\text{c}}$ and the intersect $D_1$. The dotted lines symbolize upper and lower limits (see table~\ref{tab:new_Arrhenius_parameters}).}
\label{fig:Subplots}
\end{figure}
Figure~\ref{fig:Subplots}a shows effective hydrogen diffusivities measured and calculated in other studies, as a function of the inverse temperature. From these data, the corresponding Arrhenius parameters were calculated. In figure~\ref{fig:Subplots}b, the pre-factors are plotted on a logarithmic scale, as a function of the activation energies. It can be observed, that there is a linear correlation between the two parameters. This correlation can be described by 
\begin{equation}
\label{equ:Arrhenius2}
D_0=D_{1} e^{\frac{E_A}{E_\text{c}}}\text{,}
\end{equation}
where $D_{1}$ is a new pre-factor and $E_\text{c}$ is a characteristic energy that describes this correlation. Inserting equation~\ref{equ:Arrhenius2} into the Arrhenius equation (equation~\ref{equ:Arrhenius}) yields 
\begin{equation}
\label{equ:Arrhenius_new}
D=D_{1} e^{\frac{E_A}{E_\text{c}}} e^{-\frac{E_A}{k_BT}}=D_{1}e^{E_A\frac{k_BT-E_\text{c}}{E_\text{c}k_BT}}\text{,}
\end{equation}
 These new parameters can be found in table~\ref{tab:new_Arrhenius_parameters}.
\begin{table}
\begin{ruledtabular}
\begin{tabular}{c c c c}
& Best fit& Upper limit & Lower limit \\
  \colrule
$E_\text{c} [eV]$ & {0.075} & {0.089} & {0.064}\\
$D_1$ [$\text{cm}^{2}\text{s}^{-1}$] & {$1.5 \times 10^{-7}$} & {$1.7 \times 10^{-4}$} &{$1.4 \times 10^{-10}$}\\
$T_\text{c}$ [$^{\circ}$C]& {594}&{765}&{472} \\
\end{tabular}
\end{ruledtabular}
\caption
{\label{tab:new_Arrhenius_parameters}
Characteristic energy $E_\text{c}$ and new pre-factor $D_1$ according to equation~\ref{equ:Arrhenius2} and characteristic temperature $T_\text{c}$.}
\end{table}

In figure~\ref{fig:Subplots}a the best fits were used to plot the dashed lines by calculating the temperature dependence of the diffusivity for activation energies of \eV{0.5} and \eV{3}. The upper and lower limits were used to plot the dotted lines. Although our measured Arrhenius parameters ($E_A$ and $D_0$) are higher than most reported in the literature, they nicely fit this trend.

The characteristic energy $E_\textrm{c}$ equals $k_BT$ at the temperature $T_\textrm{c}$ which is at \Celsius{594} for the best fit. At this temperature the hydrogen diffusivity equals $D_1$ and is the same for different activation energies.

\subsection{\label{sec:Influences}Influences on the Effective Diffusivity of Hydrogen in Silicon}

The effective diffusivity of hydrogen in silicon is influenced by various material properties and other parameters which differ between the different studies shown in figure~\ref{fig:Subplots}. To determine which materials properties have the strongest influence on the hydrogen diffusivity, we grouped the diffusivity data according to the crystallization technique, the doping type and the method of introducing hydrogen. 

Various crystallization techniques have been used in hydrogen diffusion experiments. In this study the substrate materials were magnetic Czochralski (m:Cz) and float zone (FZ) wafers. FZ material was also used in references~\citenum{Ichimiya1968,Rizk1991,Johnson1991,Sopori1992} amd \citenum{Laven2013}. Other groups also investigated conventional Czochralski (Cz) pulled silicon \cite{vanWieringen1956,Mogro-Campero1985,Newman1991,Johnson1991,Sopori1992,Huang2004}. The different materials roughly differ in their impurity concentrations. Cz has the highest concentrations of oxygen and carbon. These concentrations are smaller in m:Cz and even smaller in FZ-silicon. For a comparison also some studies on the effective hydrogen diffusivity in poly-crystalline silicon (poly) are included\cite{Kazmerski1985,Sopori1992}.

Another possible influence on the effective diffusivity of hydrogen is the doping type and the concentration of dopands. Within the same study, the effective diffusivity of H$^+$ which is the dominant hydrogen species in $p$-type material, is higher than the diffusivity of H$^-$ in $n$-type material\cite{Rizk1991}.

Also the method of introducing the hydrogen into the sample differs between different studies. The hydrogen can be implanted as protons at high \cite{Laven2013} and at low \cite{Kazmerski1985,Seager1988} energies or using a hydrogen plasma \cite{Mogro-Campero1985,Pearton1985,Herrero1990,Rizk1991,Johnson1991,Sopori1992,Huang2004}. Another way of inserting hydrogen is by using permeation \cite{vanWieringen1956,Ichimiya1968}. Using proton implantation most of the hydrogen is deposited in a very narrow region around the implantation depth. This depth depends on the implantation energy. While at an implantation energy of \eV{100} the implantation depth is around \nm{4}, this depth is increased to around \mum{70} at an energy of \MeV{2.5}. The concentration of hydrogen at the implantation depth scales inversely with the implantation energy. After implanting the same dose of protons, the concentration of hydrogen at the implantation depth is almost three orders of magnitude higher if an implantation energy of \eV{100} is used compared to an energy of \MeV{2.5}. During a plasma treatment, the highest concentration of hydrogen is at the surface of the material and is usually higher than \concentration{$10^{19}$}. Proton implantation also changes the silicon lattice as it generates high concentrations of vacancies and interstitials. This most certainly also influences the effective diffusivity of hydrogen through this region towards the surface. Apart from gas permeation, all other techniques to introduce hydrogen yield maximum concentrations which are far higher than the solubility limit of hydrogen. At the investigated temperatures, the solubility of hydrogen lies between \concentration{$1*10^8$} and \concentration{$5*10^{11}$} (see references \citenum{vanWieringen1956} and \citenum{Binns1993}).

Some studies, which are also included in figure~\ref{fig:Subplots} focused on the simulation of the hydrogen diffusivity using diffusion-reaction\cite{Capizzi1987} or molecular dynamics\cite{Panzarini1994,Buda1989} simulations.
\begin{figure*}[bpt]%
 \includegraphics[width=\linewidth]{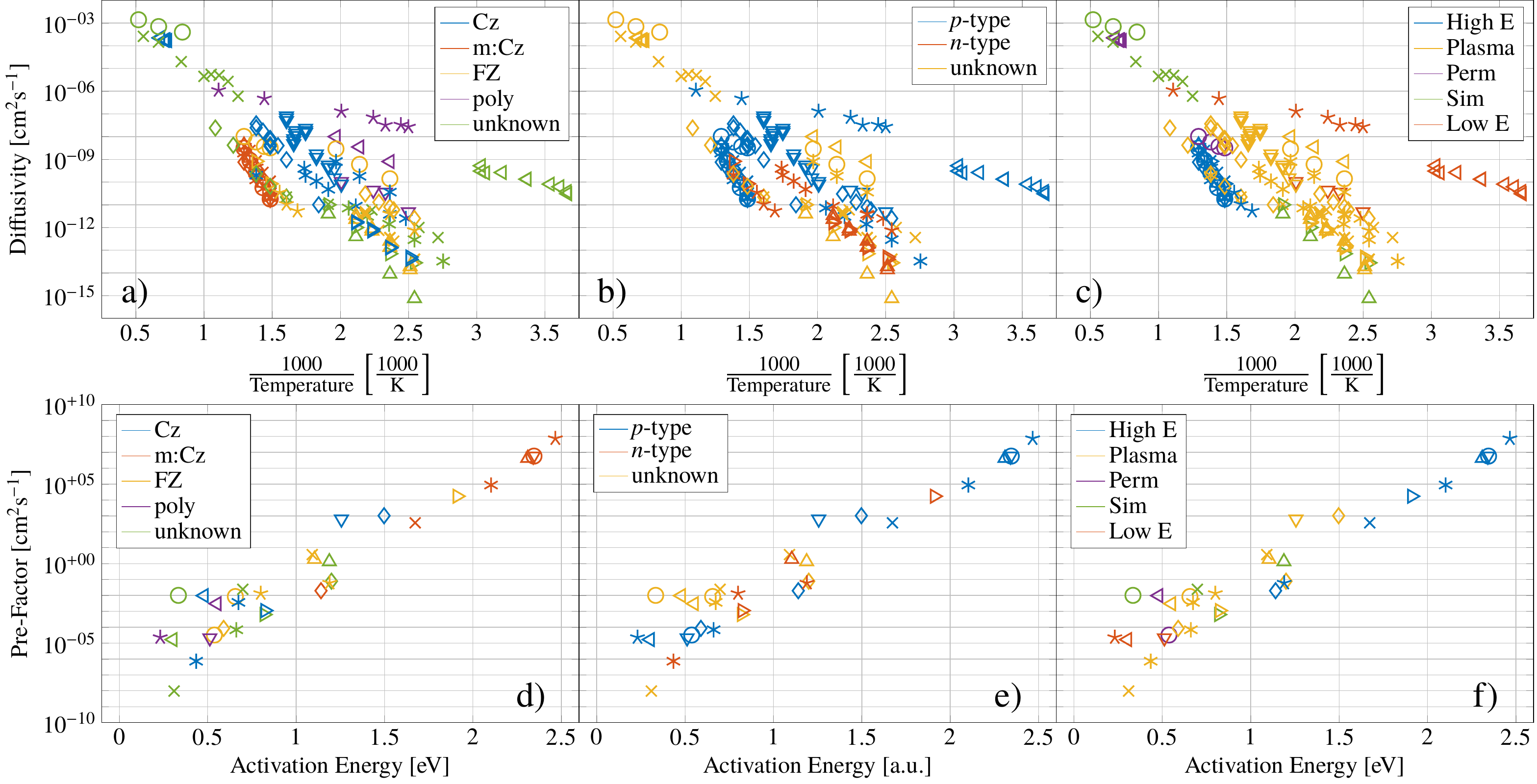}
\caption{Influence of different parameters on the effective hydrogen diffusivity as a function of the inverse annealing temperature (a-c) and on the Arrhenius coefficients of the hydrogen diffusion (d-f). a,d:~Crystallization technique. b,e:~Doping type. c,f:~Method of inserting hydrogen (High E\--high energy ion implantation, Perm\--permeation, Sim\--simulation, Low E\--low energy ion implantation). }
\label{fig:Diffusivities_subfigures}
\end{figure*}

An overview on the impact of certain parameters on the hydrogen diffusivity and the corresponding Arrhenius coefficients can be found in figures~\ref{fig:Diffusivities_subfigures}. The influence of the crystal quality of the substrate material on the hydrogen diffusivity is illustrated in figure~\ref{fig:Diffusivities_subfigures}a. The plot shows that the effective hydrogen diffusivity is smaller in m:Cz-material than in Cz or FZ. The highest Arrhenius coefficients (see figure~\ref{fig:Diffusivities_subfigures}d) are found in m:Cz material while smaller coefficients were calculated for FZ- and Cz-silicon. 

The influence of the doping type and hence the diffusing species (see figures~\ref{fig:Diffusivities_subfigures}b and~\ref{fig:Diffusivities_subfigures}e) on the hydrogen diffusivity does not seem to be of great relevance when different studies are compared. This could be because the Fermi energy moves to the middle of the band gap during high temperature anneals making all samples intrinsic during the anneal. 

The method of introducing the hydrogen has a strong influence on the diffusivity of hydrogen (see figures~\ref{fig:Diffusivities_subfigures}c and~\ref{fig:Diffusivities_subfigures}f). The comparison of the different studies showed that the highest diffusivities associated with low Arrhenius coefficients resulted after low energy proton implantation as well as after plasma treatments. The lowest values for the hydrogen diffusivity and the highest Arrhenius coefficients were observed after high energy proton implantation. 

\subsection{\label{sec:HConc}Dependence of the Effective Hydrogen Diffusivity on the Hydrogen Concentration}

\begin{figure}[tp]%
 \includegraphics[width=\linewidth]{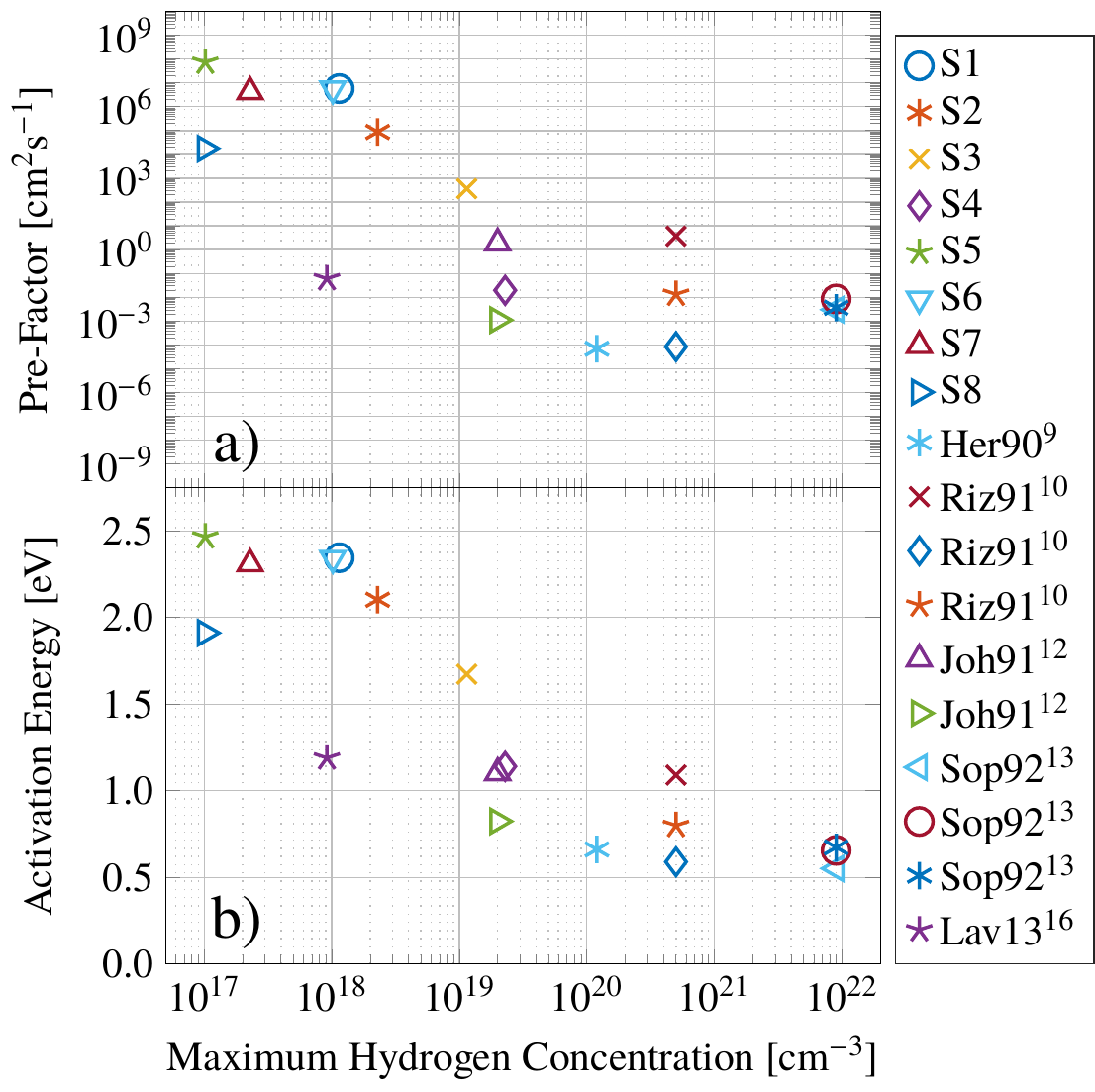}
\caption{Arrhenius parameters ( a: D$_0$, b: E$_\textrm{A}$ ) as a function of the maximum concentration of hydrogen in the sample.}
\label{fig:Ea_Hconc_literature}
\end{figure}

Both, the dependence of the Arrhenius parameters on the proton implantation dose (as shown in figure~\ref{fig:Arrhenius_parameters}), as well as their dependence on the method of hydrogen introduction (see figure~\ref{fig:Diffusivities_subfigures}f) suggest a correlation of the activation energy and the hydrogen concentration. The correlation of the activation energy to the maximum hydrogen concentration is shown in figure~\ref{fig:Ea_Hconc_literature}.  
Here, the activation energy is plotted as a function of the logarithm of the maximum hydrogen concentration. In this study and in reference~\citenum{Laven2013}, proton implantation was used to introduce the hydrogen. For proton implanted samples, the maximum hydrogen concentration can be calculated from the implantation energy and the implanted proton dose using hydrogen concentration profiles simulated with SRIM\cite{SRIM2013,Ziegler2010}. Other concentration values were extracted from concentration profiles of hydrogen defects measured with infra red-reflectance (IR) \cite{Herrero1990}, or from concentration profiles of hydrogen or deuterium measured using secondary ion mass spectroscopy (SIMS)\cite{Rizk1991,Johnson1991,Sopori1992}.

To understand why the activation energy correlates to the hydrogen concentration, consider where the hydrogen will reside in the crystal. Hydrogen preferentially forms defect complexes with the defect that minimizes the energy. When the concentration of hydrogen is lower than the concentration of these defects, the diffusion of hydrogen is limited by the speed of reactions with these defects and dissociations from them. The activation energy for this kind of movement at low hydrogen concentrations $E_{lo}$ is related to the average binding energy of hydrogen to these preferred defects. The preferred defect will depend on how the crystal was prepared. Binding energies of hydrogen to impurities in silicon can range between less than \eV{0.5} and more than \eV{2.5} \cite{VandeWalle1994,Chang1988,Bergman1988,Zhu1990}. If the concentration of hydrogen is increased, hydrogen atoms will saturate the preferred defect and will start occupying lattice sites of higher energy. Hence, the average energy barrier for migration decreases. Once the hydrogen concentration is far higher than the concentration of all defects combined, defect levels are saturated and the activation energy for the diffusion of hydrogen converges to $E_{hi}$, an energy, related the barrier between two lattice sites. 

The correlation of the activation energy and the maximum hydrogen concentration is described by the following set of equations:
\begin{equation}
\label{equ:Activation_Energy}
\begin{aligned}
	\text{[H]}<\text{[H]}_{E_{lo}} &\rightarrow E_A=E_{lo} \\
	[\text{H}]_{E_{lo}}<\text{[H]}<[\text{H}]_{E_{hi}} &\rightarrow E_A=E_{lo}-E_T \text{log} \left( \frac{[\text{H}]}{[\text{H}]_{E_{lo}}}\right) \\
	\text{[H]}>[\text{H}]_{E_{hi}} &\rightarrow E_A=E_{hi} 
\end{aligned}
\end{equation}
Here, [H]$_{E_{lo}}$ is the critical hydrogen concentration below which the activation energy $E_A$ equals $E_{lo}$, an energy related to the mean binding energy of hydrogen to impurities. [H]$_{E_{hi}}$ is the concentration of hydrogen above which the activation energy equals $E_{hi}$, an energy associated to the migration of hydrogen without interacting with impurities. $E_T$ is the transition energy, associated with the decay of the activation energy with increasing hydrogen concentration between [H]$_{E_{lo}}$ and [H]$_{E_{hi}}$. 

\begin{figure*}[tp]%
 \includegraphics[width=\linewidth]{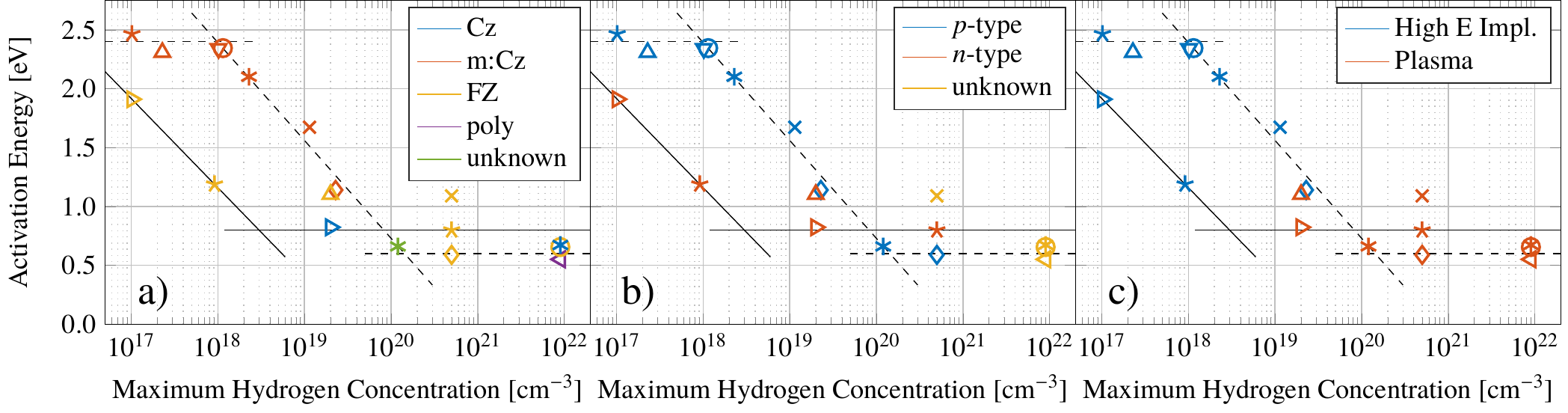}
\caption{Influence of different parameters on the activation energy as a function of the maximum concentration of hydrogen in the sample. a:~Crystallization technique. b:~Doping type. c:~Method of inserting hydrogen (High E - high energy ion implantation). Solid line: fit for $n$-type FZ. Dashed line: fit for $p$-type m:Cz. }
\label{fig:Ea_Hconc_subfigures}
\end{figure*}

The critical hydrogen concentrations, as well $E_{hi}$, $E_{lo}$ and $E_{T}$ are material dependent. In figure~\ref{fig:Ea_Hconc_subfigures} the influences of the crystallization technique, the doping type and the method of hydrogen introduction on the activation energy are plotted once more. 

Regions of constant activation energy and regions which show a linear dependence of the activation energy on the logarithmic hydrogen concentration are indicated by solid and dashed lines. The solid lines represent fits through $p$-type m:Cz data points and the dashed lines fit the $n$-type FZ data. 
In $p$-type m:Cz material the activation energy $E_{lo}$=\eV{2.4} at hydrogen concentrations smaller than $[$H$]_{E_{lo}}$=\concentration{$10^{18}$}. $E_A$ then decays linearly between $[$H$]_{E_{lo}}$ and $[$H$]_{E_{hi}}$=\concentration{$10^{20}$} following a transition energy of $E_T$=\eV{0.83} and equals an energy $E_{hi}$ of \eV{0.6} at hydrogen concentrations higher than $[$H$]_{E_{hi}}$. In $n$-type FZ material a transition energy of \eV{0.76} was observed. At hydrogen concentrations above $[$H$]_{E_{hi}}$=\concentration{$10^{18}$}, the activation energy equals $E_{hi}$=\eV{0.8}. In the case of $p$-type m:Cz material $E_T$ is approximately ($E_{lo}$-$E_{hi}$)/2. Using the same relation for $n$-type FZ material, $E_{lo}$ and [H]$_{E_{lo}}$ can be approximated to \eV{2.3} and \concentration{$3\times 10^{16}$}. In table \ref{tab:Activation_Energy} the parameters describing the dependence of the activation energy on the hydrogen concentration in $p$-type m:Cz and in $n$-type FZ material are listed.
\begin{table}[tp]%
\begin{ruledtabular}
\begin{tabular}{c c c }
	&\emph{p}-type m:Cz&\emph{n}-type FZ\\
    \colrule
    $E_{lo}$ [eV] & 2.4& $2.3^*$\\
    $E_T$ [eV] & ${0.83}$ & ${0.76}$ \\
    $E_{hi}$ [eV]& 0.6  & 0.8\\
     $[$H$]_{E_{lo}}$ [cm$^{-3}$]& $10^{18}$ & $3\times{10^{16}}^*$\\
    $[$H$]_{E_{hi}}$ [cm$^{-3}$]& $10^{20}$ & $3 \times 10^{18}$\\
\end{tabular}
\end{ruledtabular}
\caption
{\label{tab:Activation_Energy} Parameters describing the dependence of the activation energy on the hydrogen concentration. $^*$Estimated values for $n$-type float zone at low hydrogen concentrations.
}
\end{table}


The investigation of the different hydrogen introduction methods (figure \ref{fig:Ea_Hconc_subfigures}c) showed that low hydrogen concentrations where achieved, when high energy proton implantation was used. In such samples the hydrogen concentration was always measured using SRP. High concentrations of hydrogen associated with a small activation energy where observed after plasma treatments. Here, SIMS or IR were used to measure the hydrogen concentration. 


\section{\label{sec:Conclusion}Conclusion}
In this study, we investigated the effective diffusion of hydrogen depending on the implantation dose in high energy proton implanted silicon using SRP. We present a model where hydrogen diffuses by moving from defect to defect in the silicon. Hydrogen attaches to some defect and then is released in a thermally activated process to later attach at some other defect. This can be used to explain the wide variation in observed diffusion constants for hydrogen in silicon. The diffusion depends on the type and the concentration of defects present. This model also explains the observed hydrogen concentration dependence on the diffusion. Hydrogen first attaches to the defects which allows it to lower its energy the most. As these defects become saturated with hydrogen, the hydrogen occupies sites at defects that are energetically less favorable. Since the dissociation from the defects is thermally activated, this leads to a faster diffusion of hydrogen for higher concentrations. By comparing our work to other studies on the diffusion of hydrogen, a correlation between the activation energy $E_A$ and the pre-factor $D_0$ has been observed, $D_0 = D_1\exp (E_A/E_c)$, where $D_1$ is~\diffusivity{$1.5 \times 10^{-7}$} and the characteristic energy $E_\textrm{c}$ is~\eV{0.075}. This correlation is associated with a temperature $T_c$ (\Celsius{594}), at which the effective diffusivity equals $D_1$ and is the same for different activation energies. Furthermore, the comparison of the different studies showed a correlation of the activation energy to the maximum hydrogen concentration (Eq. 6). This correlation is different for materials containing different impurity concentrations such as $n$-type FZ and $p$-type m:Cz material. It is characterized by a constant activation energy $E_{lo}$ at low hydrogen concentrations, attributed to an average binding energy to impurities. At high hydrogen concentrations the activation energy $E_{hi}$ is also constant and might be associated with the energy barrier between two lattice sites . At intermediate hydrogen concentrations a transition from $E_{lo}$ to $E_{hi}$ has been observed which is a linear function of the logarithm of the maximum hydrogen concentration. The slope of this transition is described by the transition energy $E_T$ which is approximately the average of $E_{lo}$ and $E_{hi}$. The results show that for the same hydrogen concentration the activation energy for the hydrogen diffusion is smaller in FZ than in Cz silicon due to the smaller concentration of impurities in the material.

\begin{acknowledgments}
This work has been performed in the project EPPL, co-funded by grants from Austria, Germany, The Netherlands, Italy, France, Portugal- ENIAC member States and the ENIAC Joint Undertaking. This project is co-funded within the program "Forschung, Innovation und Technologie f\"{u}r Informationstechnologie" by the Austrian Ministry for Transport, Innovation and Technology.
\end{acknowledgments}

\bibliography{literature}

\end{document}